\documentclass[copyright,creativecommons]{eptcs}
\providecommand{\event}{UITP 2012}
\usepackage{breakurl}
\usepackage{graphicx}

\usepackage{ifthen}
\usepackage{isabelle,isabellesym}

\isadroptag{theory}
\isabellestyle{it}

\newcommand{\secref}[1]{\S\ref{#1}}

\newcommand{\isasymPR}{\isacommand{pr}}
\newcommand{\isasymNITPICK}{\isacommand{nitpick}}
\newcommand{\isasymQUICKCHECK}{\isacommand{quickcheck}}
\newcommand{\isasymSLEDGEHAMMER}{\isacommand{sledgehammer}}

\newcommand{\isasymINDUCTIVE}{\isacommand{inductive}}

\newcommand{\isasymLET}{\isacommand{let}}
\newcommand{\isasymIN}{\isacommand{in}}

\newcommand{\isasymUNDO}{\isacommand{undo}}
\newcommand{\isasymREDO}{\isacommand{redo}}
\newcommand{\isasymSORRY}{$\;\langle\mathit{proof}\rangle$}

\renewcommand{\isacommand}[1]
{\ifthenelse{\equal{sorry}{#1}}{\isasymSORRY}{\isakeyword{#1}}}

\newcommand{\para}[1]{\medskip\par\noindent\textbf{#1}}

\begin{document}

\title{READ-EVAL-PRINT in \\ Parallel and Asynchronous Proof-checking}
\def\titlerunning{READ-EVAL-PRINT in Parallel and Asynchronous Proof-checking}

\author{Makarius Wenzel \thanks{Current research supported by Digiteo Foundation
  and Project Paral-ITP (ANR-11-INSE-001).} \\
  \institute{Univ. Paris-Sud, Laboratoire LRI, UMR8623, Orsay, F-91405, France \\
  CNRS, Orsay, F-91405, France}}
\def\authorrunning{M. Wenzel}
\maketitle

\begin{abstract}
  The LCF tradition of interactive theorem proving, which was started
  by Milner in the 1970-ies, appears to be tied to the classic
  READ-EVAL-PRINT-LOOP of sequential and synchronous evaluation of
  prover commands.  We break up this loop and retrofit the
  read-eval-print phases into a model of parallel and asynchronous
  proof processing.  Thus we explain some key concepts of the
  Isabelle/Scala approach to prover interaction and integration, and
  the Isabelle/jEdit Prover IDE as front-end technology.  We hope to
  open up the scientific discussion about non-trivial interaction
  models for ITP systems again, and help getting other old-school
  proof assistants on a similar track.
\end{abstract}

\begin{isabellebody}%
\def\isabellecontext{Paper}%
\isadelimtheory
\endisadelimtheory
\isatagtheory
\isacommand{theory}\isamarkupfalse%
\ Paper\isanewline
\isakeyword{imports}\ Base\isanewline
\isakeyword{begin}%
\endisatagtheory
{\isafoldtheory}%
\isadelimtheory
\endisadelimtheory
\isamarkupsection{Introduction%
}
\isamarkuptrue%
\isamarkupsubsection{Motivation%
}
\isamarkuptrue%
\begin{isamarkuptext}%
Isabelle \cite[\S6]{Wiedijk:2006} is one of the classic
  members of the LCF prover family, together with Coq
  \cite[\S4]{Wiedijk:2006} and the variety of HOL systems
  \cite[\S1]{Wiedijk:2006}.  The survey on Isabelle
  \cite{Wenzel-Paulson-Nipkow:2008} from 2008 provides some entry
  points to the diverse tools, packages, and applications of our
  prover platform.  It started as a pure logical framework in 1989 and
  has grown into a general framework for integrating logic-based
  tools, including automated provers and disprovers.  The
  Isabelle2008 version also marks the turning point of substantial
  reforms in the organization of the proof process, such that it works
  efficiently on multi-core hardware, which is now commonplace.

  The original work on parallel Poly/ML and Isabelle/ML is reported in
  \cite{Matthews-Wenzel:2010,Wenzel:2009,Wenzel:2013:ITP}.  The core
  idea is to provide a parallel LCF-style inference kernel that
  supports a concept of \emph{proof promises} natively, and to
  integrate it with the task-parallel library for \emph{future values}
  in Isabelle/ML.  The general principle behind this is \emph{managed
  evaluation} in ML: the system organizes the execution of user code,
  similar to an operating system that runs user processes.  Managed
  evaluation includes external POSIX shell processes run from
  Isabelle/ML, which is used in Sledgehammer to run provers
  implemented in C or to access the \emph{System on TPTP} service at
  \url{http://www.cs.miami.edu/~tptp/cgi-bin/SystemOnTPTP}.

  \medskip Soon after the initial success of parallel Isabelle it
  became clear that ambitious forking of proofs is in conflict with
  the received interaction model of the TTY loop, and its canonical
  front-end Proof~General \cite{Aspinall:TACAS:2000}.  To illustrate
  this, we consider the following example:%
\end{isamarkuptext}%
\isamarkuptrue%
\isacommand{inductive}\isamarkupfalse%
\ path\ \isakeyword{for}\ R\ {\isacharcolon}{\isacharcolon}\ {\isachardoublequoteopen}{\isacharprime}a\ {\isasymRightarrow}\ {\isacharprime}a\ {\isasymRightarrow}\ bool{\isachardoublequoteclose}\ \ %
\isamarkupcmt{implicit proofs: monotonicity and derived rules%
}
\isanewline
\isakeyword{where}\isanewline
\ \ base{\isacharcolon}\ {\isachardoublequoteopen}path\ R\ x\isactrlisub {\isadigit{1}}\ x\isactrlisub {\isadigit{1}}{\isachardoublequoteclose}\isanewline
{\isacharbar}\ step{\isacharcolon}\ {\isachardoublequoteopen}R\ x\isactrlisub {\isadigit{1}}\ x\isactrlisub {\isadigit{2}}\ {\isasymLongrightarrow}\ path\ R\ x\isactrlisub {\isadigit{2}}\ x\isactrlisub {\isadigit{3}}\ {\isasymLongrightarrow}\ path\ R\ x\isactrlisub {\isadigit{1}}\ x\isactrlisub {\isadigit{3}}{\isachardoublequoteclose}\isanewline
\isanewline
\isacommand{theorem}\isamarkupfalse%
\ example{\isacharcolon}\ {\isachardoublequoteopen}path\ R\ x\isactrlisub {\isadigit{1}}\ x\isactrlisub {\isadigit{3}}\ {\isasymLongrightarrow}\ P\ x\isactrlisub {\isadigit{1}}\ x\isactrlisub {\isadigit{3}}{\isachardoublequoteclose}\isanewline
\isadelimproof
\endisadelimproof
\isatagproof
\isacommand{proof}\isamarkupfalse%
\ {\isacharparenleft}induct\ rule{\isacharcolon}\ path{\isachardot}induct{\isacharparenright}\ \ %
\isamarkupcmt{explicit toplevel proof%
}
\isanewline
\ \ \isacommand{case}\isamarkupfalse%
\ {\isacharparenleft}base\ x\isactrlisub {\isadigit{1}}{\isacharparenright}\isanewline
\ \ \isacommand{show}\isamarkupfalse%
\ {\isachardoublequoteopen}P\ x\isactrlisub {\isadigit{1}}\ x\isactrlisub {\isadigit{1}}{\isachardoublequoteclose}\ \isacommand{sorry}\isamarkupfalse%
\ \ %
\isamarkupcmt{explicit sub-proof%
}
\isanewline
\isacommand{next}\isamarkupfalse%
\isanewline
\ \ \isacommand{case}\isamarkupfalse%
\ {\isacharparenleft}step\ x\isactrlisub {\isadigit{1}}\ x\isactrlisub {\isadigit{2}}\ x\isactrlisub {\isadigit{3}}{\isacharparenright}\isanewline
\ \ \isacommand{note}\isamarkupfalse%
\ {\isacharbackquoteopen}R\ x\isactrlisub {\isadigit{1}}\ x\isactrlisub {\isadigit{2}}{\isacharbackquoteclose}\ \isakeyword{and}\ {\isacharbackquoteopen}path\ R\ x\isactrlisub {\isadigit{2}}\ x\isactrlisub {\isadigit{3}}{\isacharbackquoteclose}\isanewline
\ \ \isacommand{moreover}\isamarkupfalse%
\ \isacommand{note}\isamarkupfalse%
\ {\isacharbackquoteopen}P\ x\isactrlisub {\isadigit{2}}\ x\isactrlisub {\isadigit{3}}{\isacharbackquoteclose}\isanewline
\ \ \isacommand{ultimately}\isamarkupfalse%
\ \isacommand{show}\isamarkupfalse%
\ {\isachardoublequoteopen}P\ x\isactrlisub {\isadigit{1}}\ x\isactrlisub {\isadigit{3}}{\isachardoublequoteclose}\ \isacommand{sorry}\isamarkupfalse%
\ \ %
\isamarkupcmt{explicit sub-proof%
}
\isanewline
\isacommand{qed}\isamarkupfalse%
\endisatagproof
{\isafoldproof}%
\isadelimproof
\endisadelimproof
\begin{isamarkuptext}%
\noindent This formal Isar text follows the basic structure of
  mathematical documents as a sequence of definition --- statement ---
  proof, but the (inductive) definition also involves some implicit
  proofs internally.  Monotonicity of the specification is a
  prerequisite for further internal derivations of the introduction
  rules and induction principle (by the Knaster-Tarski fixed-point
  theorem).  A failure proving monotonicity indicates some mistake in
  the user specification, but a failure proving derived rules some
  problem in the tool implementation.

  The parallel batch mode of Isabelle forks all these proofs via the
  future evaluation mechanism, followed by a global join over the
  whole collection of proofs from all theories that are loaded into
  the session.  This works, because proofs are not relevant for other
  proofs to proceed: it is sufficient to ensure that ultimately all
  proofs are finished.

  \medskip This important principle of \emph{proof irrelevance}
  holds only in a weaker sense for continuous editing and continuous
  checking in interactive mode.  Some anomalies can occur if implicit
  proofs are forked blindly, because the TTY loop assumes that
  commands like \isa{{\isasymINDUCTIVE}} report synchronously about
  success or failure, before the next command is started.  This limits
  the scope of parallelism to individual command transactions: all
  local proofs would have to be joined before working on subsequent
  commands.

  Another bad effect is caused by user interrupts that interfere with
  parallel evaluation of commands.  Implicitly forked proof attempts
  that are cancelled need to be \emph{restarted}.  Otherwise it could
  happen that the derivation of theorem \isa{example} above might
  contain memo-ized interrupt exceptions in the justification for the
  \isa{path{\isachardot}induct} rule.\footnote{The deeper problem is the
  non-monotonic behavior of future cancellation: a parallel evaluation
  that is not yet consolidated and cancelled cannot be continued
  afterwards.  This does not happen in batch mode, because
  cancellation means to terminate the whole process (after printing
  all failures encountered so far).  In interactive mode, the user
  expects to be able to continue editing after each round of
  cancellation caused by the incremental editing model.}

  Apart from these problems of implicit proofs in seemingly atomic
  commands, parallel processing of explicit proofs given as separate
  command sequences in the text is even further removed from the
  traditional interaction model of step-wise proof scripting.  The
  rich structure of proof texts --- with its potential for forking
  validations of proofs and processing sub-proofs independently --- is
  flattened according to depth-first traversal in classic proof
  scripting.

  \medskip These observations should make sufficiently clear that the
  classic REPL concepts require substantial reforms, to make them fit
  for the combination of asynchronous interaction with parallel proof
  processing.

  These investigations have already started in summer 2008, but it has
  required several years to get to reasonably robust implementations
  in Isabelle/Scala and Isabelle/jEdit.  An early version is outlined
  in \cite{Wenzel:2010}, the first stable release of Isabelle2011-1
  (October 2011) is presented in \cite{Wenzel:2012:CICM}.  This
  infrastructure for continuous proof checking and Prover IDE support
  is consolidated further in Isabelle2012 (May 2012) and Isabelle2013
  (February 2013), but many of the underlying concepts still need to
  be communicated. The present paper is a further step towards that.%
\end{isamarkuptext}%
\isamarkuptrue%
\isamarkupsubsection{Classic REPL Architecture \label{sec:REPL}%
}
\isamarkuptrue%
\begin{isamarkuptext}%
The classic READ-EVAL-PRINT-LOOP is well-known from
  long-standing LISP tradition.  From there it made its way into
  applications of symbolic computation, computer algebra, interactive
  theorem proving etc.  The basic idea is to process a sequence of
  \emph{commands} one by one, and report results immediately to the
  user.

  The division into the main phases of the loop can be explained in a
  first approximation from the perspective of the LISP interpreter,
  which processes a sequence of LISP expressions as toplevel
  declarations as follows.

  \begin{description}

  \item[READ:] process the syntax of the given expression ---
  \emph{internalize} it as semantic operation on the program state.

  \item[EVAL:] evaluate the internalized expression in the current
  state --- \emph{run} it and update the toplevel state accordingly.

  \item[PRINT:] output the result of the evaluation ---
  \emph{externalize} values, usually in the same notation as the
  input.

  \item[LOOP:] continue the above \emph{ad infinitum} (or until the user
  terminates the command interpreter).

  \end{description}

  The READ-EVAL-PRINT phases structure various situations within the
  interpreter, and the LOOP phase defines the interactive behavior of
  the system.  The latter involves some technical details about
  organizing interaction that are often taken for granted in the
  folklore history of these concepts.  Subsequently we recall some of
  this common ground and relate it to issues faced by classic REPL
  front-ends like Proof~General \cite{Aspinall:TACAS:2000} and refined
  versions of its protocols in PGIP \cite{Aspinall-et-al:2007}.

  \para{Prompt.} The system prints a command prompt and flushes the
  output channel to ensure the user can see it, and awaits
  input.\footnote{Flushing is sometimes forgotten in implementations
  and only discovered when the system is run over a pipe for the first
  time, without the automatic per-line flush of the TTY stream on
  Unix.}

  Conceptually, the prompt behavior means full synchronization of the
  pair of input/output channels.  This incurs certain real-time
  delays, say in local interprocess-communication to flush the buffers
  of the connecting pipe.  For network connections the extra latency
  of a full round-trip needs to be taken into account.  This does not
  prevent implementation of distributed editors on the World Wide Web
  such as Etherpad \url{http://etherpad.com}, but the throughput of
  such synchronized interaction is limited.

  Proof~General uses the command prompt as the main protocol marker
  --- the prover is required to decorate its prompt by special control
  sequences.  This allows to separate command boundaries semantically:
  all observable output from the evaluation phase between two command
  prompts is attached to the corresponding command span in the source
  text.  This natural observation of the TTY loop imposes some
  limitations on command evaluation strategies, though.  It is
  difficult to detach asynchronous commands from the main loop ---
  deferred output can confuse processing of other commands.  The user
  needs to understand the meaning of displaced messages, and
  occasionally ``repair'' the protocol by issuing suitable control
  commands for re-synchronization of the editor with the prover.

  \para{Handling of errors.} Any of the READ-EVAL-PRINT phases might
  fail, which results in some error output instead of regular PRINT.
  The LOOP needs to ensure that command transactions are atomic: the
  toplevel state is only updated after a successful run; errors should
  result in a clean \emph{rollback} to the previous state.  This
  means, a failing command transaction essentially results in an
  identity function on the state with some extra output, but it
  depends on details of the prover if it moves one step forward in the
  command execution, or not.  This might affect subsequent navigation
  of the command history (\emph{undo}).

  Classic Proof~General and especially PGIP attempt to formalize such
  notions of ``success'' or ``error'' of command transactions, such
  that both the editor and the prover agree on it (at least in
  theory).  This still poses problems in boundary cases, with
  debatable situations of \emph{non-fatal} errors that look like a
  command failure, but are intended as a strong warning issued by a
  successful command.  It also explains why developers of Isabelle
  proof tools had to be instructed to emit error messages only if a
  subsequent failure of the whole command could be guaranteed,
  otherwise the front-end would loose synchronization.

  For robustness it is desirable to make the integrity of command
  transactions independent of accidental prover messages.  This opens
  a spectrum of informative messages, warnings, non-fatal errors,
  fatal errors etc.\ without affecting critical aspects of the
  interaction protocol.

  \para{Handling of interrupts.} The aim is to allow the user to
  intercept command execution, say by pressing CTRL-C or pushing some
  emergency brake button.  The standard implementation makes the LOOP
  itself uninterruptible, but enables interrupts for executing each
  command (especially in the EVAL phase, which might be
  non-terminating).  This assumes that the runtime environment that
  executes the command reacts accordingly and aborts the user program.

  Even many decades after the introduction of hardware interrupts and
  process signals (at least on POSIX systems), interruptibility of
  arbitrary user-code cannot be taken for granted.  Servicing of
  interrupt requests might be too slow (resulting in noticeable
  delays), or too fast (resulting in inconsistent internal program
  state).  A LISP interpreter might have no problems to poll the
  interrupt status frequently, but more advanced language platforms
  need to do more to make it work efficiently and reliably.  Poly/ML
  (which underlies Isabelle/ML) is able to handle interrupts quickly
  in most practical situations, with well-defined meaning of signals
  within a multi-threaded process.  External signals are dispatched to
  all threads that are configured to accept them, and internal signals
  are addressed to selected threads in isolation.  The JVM (which
  underlies Isabelle/Scala) follows a similar model, but is more
  reluctant to let interrupts interfere with regular user code:
  \verb|Thread.interrupt| is either serviced implicitly during
  I/O or needs to be polled explicitly via \verb|Thread.interrupted|.

  In any case, external interrupts raise delicate questions about the
  integrity of command transactions.  It depends on many
  implementation details if interrupted command transaction are
  properly rolled-back, or treated as successful without any effect.
  Adding the aspects of parallel and asynchronous execution makes it
  more difficult.  For example, detached evaluations of older commands
  might receive a signal from the current command evaluation
  unintentionally, and thus leave the front-end (and the user) in an
  unclear situation concerning the state of the prover.%
\end{isamarkuptext}%
\isamarkuptrue%
\isamarkupsubsection{Command Transactions and Document Structure \label{sec:document}%
}
\isamarkuptrue%
\begin{isamarkuptext}%
Subsequently we introduce a minimal formal model of command
  transactions and proof document structure, in order to clarify
  further elaborations of the REP model, and various required
  extensions for asynchronous interaction and parallel processing.
  The bigger picture is given by a document-oriented approach to
  prover interaction.  Its content-oriented aspects are explained in
  \cite{Wenzel:2011:CICM}.  The corresponding interaction model
  provides first-class notions of document editing with some version
  management built-in, as sketched below.  The idea is to embed
  ``small'' toplevel states into ``big'' document states, and provide
  some editing operations on that.\footnote{Strictly speaking, it is
  no longer appropriate to use the traditional term ``toplevel state''
  for the many small system configurations that are managed here
  simultaneously within the big document state.}

  \para{``Small'' toplevel state (isolated commands).} The local
  program configuration that is managed by the toplevel is represented
  as explicit value \isa{st}.  A \emph{command transaction} is
  essentially a partial function on a toplevel state: we write \isa{st\ {\isasymlongrightarrow}\isactrlsup t\isactrlsup r\ st{\isacharprime}} as relation, or \isa{st{\isacharprime}\ {\isacharequal}\ tr\ st} as function
  application.  The transaction is internally structured according to
  the classic READ-EVAL-PRINT phases.  As first approximation \isa{tr\ {\isacharequal}\ read{\isacharsemicolon}\ eval{\isacharsemicolon}\ print} is merely the sequential composition of
  certain internal operations.

  The original motivation for this sub-structuring was given by the
  LISP interpreter, with its \isa{intern}-\isa{run}-\isa{extern} phases, but our main purpose is to organize incremental
  checking of proof documents.  So we characterize the three phases by
  their relation to the toplevel state:

  \medskip
  \begin{tabular}{l}
  \isa{tr\ st\ {\isacharequal}} \\
  \quad \isa{{\isasymLET}\ x\ {\isacharequal}\ read\ src\ {\isasymIN}} \\
  \quad \isa{{\isasymLET}\ {\isacharparenleft}y{\isacharcomma}\ st{\isacharprime}{\isacharparenright}\ {\isacharequal}\ eval\ x\ st\ {\isasymIN}} \\
  \quad \isa{{\isasymLET}\ {\isacharparenleft}{\isacharparenright}\ {\isacharequal}\ print\ st{\isacharprime}\ y\ {\isasymIN}\ st{\isacharprime}} \\
  \end{tabular}
  \medskip

  This means \isa{read} is a prefix of the command transition that
  does not depend on input state, and \isa{print} a suffix that does
  not change the output state.  Only the core \isa{eval} operation
  may operate on the semantic state arbitrarily.  The \isa{src}
  input is essentially a parameter of the command transaction, i.e.\
  the concrete command span given in the text.

  In reality there might be syntax phases that do require access to
  the state, but they can be included in the inner \isa{eval}
  function.

  \para{Document structure.} The overall document structure has
  two main dimensions: \emph{local body} of text as sequence of
  commands and \emph{global outline} as directed-acyclic graph (DAG).
  The nodes of this graph may be understood as ``modules'', which are
  called ``theories'' in Isabelle, ``vernacular files'' in Coq, and
  ``articles'' in Mizar.

  In some sense this structuring of command transitions is accidental,
  but motivated by the typical situation in proof assistants:
  sequences of commands that are evaluated left-to-right and are
  organized in strictly foundational order of the theory graph.
  Cyclic module structure is not permitted, in contrast to programming
  languages like Haskell or Java.  Semantically, we can linearize the
  DAG by producing a canonical walk-through, which means a proof
  document can be considered (w.l.o.g.) as \emph{locally sequential}:

  \medskip
  \begin{tabular}{l}
  \isa{st\ {\isasymlongrightarrow}\isactrlsup t\isactrlsup r\ st{\isacharprime}\ {\isasymlongrightarrow}\isactrlsup t\isactrlsup r\isactrlsup {\isacharprime}\ st{\isacharprime}{\isacharprime}\ {\isasymdots}}
  \end{tabular}
  \medskip

  Thus we can ignore the outer DAG structure in theoretical
  considerations, although the module graph is an important starting
  point to organize the execution process efficiently in practice.
  More ambitious re-organization would take the inherent structure of
  the command sequence into account, as introduced for parallel batch
  proof-checking in \cite{Wenzel:2009,Matthews-Wenzel:2010}.

  Our reformed view on READ-EVAL-PRINT shall admit such non-trivial
  scheduling by the prover in interaction, while retaining a
  sequential reading of the text and its results that are presented to
  the user in the editor front-end.

  \para{``Big'' document state (version history).} A single
  document consists of a certain composition of command transactions
  as described above.  Document \emph{edits} can re-arrange the
  structure by inserting or removing intervals of command spans.  This
  results in different document \emph{versions} that are related by a
  certain \emph{history} of edits.  Each document version is
  implicitly associated with an \emph{execution} process that
  evaluates its content according to the original sequential reading
  of the text, but implements a certain evaluation strategy on its
  mathematical meaning, to make good use of the physical resources of
  the machine.

  The global \isa{Document{\isachardot}state} covers all these aspects, by
  providing these main operations:

  \medskip
  \begin{tabular}{l}
  \isa{Document{\isachardot}init{\isacharcolon}\ Document{\isachardot}state} \\
  \isa{Document{\isachardot}update{\isacharcolon}\ version{\isacharunderscore}id\ {\isasymrightarrow}\ version{\isacharunderscore}id\ {\isasymrightarrow}\ edit\isactrlsup {\isacharasterisk}\ {\isasymrightarrow}\ Document{\isachardot}state\ {\isasymrightarrow}\ Document{\isachardot}state} \\
  \isa{Document{\isachardot}remove{\isacharunderscore}versions{\isacharcolon}\ version{\isacharunderscore}id\isactrlbsup {\isacharasterisk}\isactrlesup \ {\isasymrightarrow}\ Document{\isachardot}state\ {\isasymrightarrow}\ Document{\isachardot}state} \
  \end{tabular}
  \medskip

  \isa{Document{\isachardot}update\ v\isactrlsub {\isadigit{1}}\ v\isactrlsub {\isadigit{2}}\ edits} updates the global document
  state by turning old \isa{v\isactrlsub {\isadigit{1}}} into new \isa{v\isactrlsub {\isadigit{2}}}, applying
  edits that operate on the command structure.  The result is
  registered within the global state.  Here identifier \isa{v\isactrlsub {\isadigit{1}}}
  needs to refer to an existing version that is reachable in the
  persistent prefix of editing history.  Identifier \isa{v\isactrlsub {\isadigit{2}}} has
  been freshly allocated by the editor, so it knows where the prover
  will continue eventually, without requiring separate communication.
  This declarative update of the document leads to modifications of
  the implicit execution process that is associated with the new
  version, re-using the partial execution state of the old one.  The
  prover determines the details according to the semantics of the
  text; the protocol refrains from speaking about that.

  Further fine points of \isa{Document{\isachardot}update} are determined by the
  structure of \isa{edit}, which is a concrete datatype with
  variants to insert or remove command spans from the text, or to
  indicate node dependencies in the DAG of modules, or to declare the
  so-called \emph{perspective} of the front-end on the document
  structure.  The latter represents the \emph{visible} parts of the
  document and thus provides hints to determine priorities for the
  incremental evaluation process: compared to a large hidden part of
  imported theory library and the potentially large unprocessed part
  of still pending text, the active area in the perspective is
  relatively small.  This locality property helps to make document
  change management reactive and scalable.

  The physical text editor is connected to the document model by
  classic GUI event handlers.  Thus various elementary editor
  operations will eventually become a sequence of document edits that
  are pipe-lined towards the prover: insert or remove text, open or
  close windows, scroll within open windows etc.  The granularity of
  document versions is determined implicitly via real-time delays (in
  the range of 50--500\,ms), such that edits are grouped and not every
  keystroke is passed through the interaction protocol.

  \medskip \isa{Document{\isachardot}remove{\isacharunderscore}versions} informs the prover that
  the editor is no longer interested in recent parts of the history;
  this amounts to de-allocation of resources in the document model.
  In practice it is sufficient to keep a short prefix of the editing
  history persistent, one that is able to cover the distance of
  physical editor buffer state, the document versions that are
  processed in the pipeline towards the prover, and the actual
  execution process that is currently run by the prover.  The
  Isabelle2013 implementation prunes the history periodically every
  60\,s.%
\end{isamarkuptext}%
\isamarkuptrue%
\isamarkupsection{READ-EVAL-PRINT revisited \label{sec:async-repl}%
}
\isamarkuptrue%
\isamarkupsubsection{Prover Syntax (READ) \label{sec:read}%
}
\isamarkuptrue%
\begin{isamarkuptext}%
Prover syntax is a surprisingly difficult topic, especially in
  Isabelle with its many layers, several of them with computationally
  complete mechanisms to operate on user input: syntax translations,
  type-reconstruction in multiple stages etc.  A general approach to
  reform LCF-style provers to reveal some aspects of their internal
  semantic content is explained in \cite{Wenzel:2011:CICM}.

  For the present purpose of prover interaction, it is sufficient to
  consider the superficial command language, which is called
  \emph{outer syntax} in Isabelle/Isar, and \emph{vernacular} in Coq.
  This means we need to cover only the first two layers of Isabelle
  syntax, and ignore the other 10 or so.

  Historically, the Isar language was designed at the same time as
  early versions of Proof~General, which explains some syntactic
  details of the language that allow a modest Emacs LISP program to
  discover so-called ``command-spans'' reliably in the text.  Thus
  users need to write quotes around the ``inner syntax'' of the
  logical framework, but this enables simple and robust separation of
  command boundaries.  In contrast, Proof~General for Coq involves a
  few more heuristics and approximations.

  Despite such simplifications, the cumulative CPU resources for
  parsing command spans as the user is editing the text can approach
  the same order of magnitude as proof checking itself.  In typical
  applications, only few proof commands consume significant evaluation
  time, but many commands require a certain overhead for the concrete
  syntax.

  \medskip In \secref{sec:document} we have already isolated the
  \isa{read} phase of the command transaction as a part that is
  independent of the semantic state.  This means we can reorganize the
  command application sequence to perform all \isa{read} phases
  independently, before starting to evaluate the composition:

  \medskip
  \begin{tabular}{llllll}
  \isa{{\isasymdown}\isactrlsup r\isactrlsup e\isactrlsup a\isactrlsup d}  &                  & \isa{{\isasymdown}\isactrlsup r\isactrlsup e\isactrlsup a\isactrlsup d}   & & \isa{{\isasymdots}} \\
  \isa{st}    & \isa{{\isasymlongrightarrow}\isactrlsup e\isactrlsup v\isactrlsup a\isactrlisup l} & \isa{st{\isacharprime}}    & \isa{{\isasymlongrightarrow}\isactrlsup e\isactrlsup v\isactrlsup a\isactrlsup l} & \isa{st{\isacharprime}{\isacharprime}} & \isa{{\isasymdots}} \\
  \end{tabular}
  \medskip

  \noindent The \isa{read} phase is required to be a total operation
  that terminates quickly.  Syntax errors need to be encoded into the
  result, e.g.\ by producing error tokens, and postponing actual
  runtime exceptions to the \isa{eval} phase that runs the
  internalized command text later on.

  Nonetheless, the result of the preliminary \isa{read} phase
  already contains useful information about the basic structure of the
  text, such as keywords and quoted text ranges that may be reported
  back to the front-end to produce some syntax-highlighting, based on
  authentic information from the prover, not the approximations as
  regular-expressions that are often seen in editors.

  \medskip The diagram above admits at least two further
  re-organizations to improve performance.

  \para{(1) Internalization} of results of each \isa{read} of the
  command source, such that it can be referenced later by some
  symbolic \isa{id} (notably in operations of the document model).
  To achieve this we provide an auxiliary operation on the ``big''
  document state:

  \medskip
  \begin{tabular}{l}
  \isa{Document{\isachardot}define{\isacharunderscore}command{\isacharcolon}\ id\ {\isasymrightarrow}\ string\ {\isasymrightarrow}\ string\ {\isasymrightarrow}\ Document{\isachardot}state\ {\isasymrightarrow}\ Document{\isachardot}state} \
  \end{tabular}
  \medskip

  \noindent \isa{Document{\isachardot}define{\isacharunderscore}command\ id\ name\ src} registers
  some command \isa{src} text for further use via \isa{id}.  The
  \isa{name} is an aspect of the parsed content that has already
  been discovered by approximative parsing in the editor; it helps the
  prover to organize document processing before commencing the actual
  \isa{read} phase.

  \medskip In Isabelle2012 and Isabelle2013 \isa{read} means to scan
  Isabelle ``symbols'' (ASCII + UTF8 text characters + infinitely many
  named mathematical symbols like \verb,\<forall>,), and to tokenize according
  to outer syntax keyword tables and some fixed formats for
  identifiers and quoted text ranges.

  In the earlier Isabelle2011-1 version, full outer syntax parsing was
  performed in the \isa{read} phase, but it now happens in \isa{eval}.  Thus we can support extension of the command language within
  Isabelle theories smoothly: for the first time in the long history
  of Isabelle, the system does not depend on external keyword tables
  generated in batch mode, and commands can be used in the same theory
  body where they are defined.  This detail is particularly important
  for developers of derivative tools in the Isabelle framework, who
  introduce their own commands, as part of regular editing and loading
  of theories that contain ML modules.

  \para{(2) Parallelization} of the \isa{read} phases, which neither
  depend on the toplevel state nor on each other.  The parsing
  involved in \isa{Document{\isachardot}define{\isacharunderscore}command} could be forked as
  future task, and joined only before command evaluation starts.

  This simple parallel parsing scheme was used in Isabelle2011-1, but
  later replaced by more modest lazy evaluation in Isabelle2012 in the
  course of some fine-tuning for machines with 2--4 cores only.  The
  reduced \isa{read} phase no longer justified the (small) overhead
  for fork/join in the preparatory stage of command transaction.  It
  might become relevant again when the system is tuned for hardware
  with 8--16 cores, where further potential for parallelism needs to
  be exploited to make use of the CPU resources.%
\end{isamarkuptext}%
\isamarkuptrue%
\isamarkupsubsection{Managed Evaluation (EVAL) \label{sec:eval}%
}
\isamarkuptrue%
\begin{isamarkuptext}%
Our standard model for evaluation of user code is that of
  Standard ML with a few restrictions and extensions.  This covers the
  following in particular:

  \begin{itemize}

  \item strict functional evaluation, without global side-effects
  (program state is managed by the value-oriented \emph{context data}
  concept of the Isabelle framework);

  \item program exceptions according to Standard ML, to indicate
  non-local exits from functional programs;

  \item physical exceptions as intrusion of the environment into the
  program execution (mapped to the special \isa{Interrupt}
  exception);

  \item potentially non-terminating execution that is always
  interruptible;

  \item I/O via official Isabelle/ML channels for \isa{writeln},
  \isa{warning}, \isa{tracing} messages etc.\ or via private
  temporary files as input to private external processes (this
  emulates value-oriented behavior on the file-system).

  \end{itemize}

  The Isabelle system infrastructure uses a variety of standard
  implementation techniques to define an explicit transaction context
  for user commands.  This includes message channels where output is
  explicitly tagged with an execution identifier, to attach it to the
  proper place in the input source, despite physical re-ordering in
  the parallel execution environment.

  Unlike a real operating system that can use hardware mechanisms to
  enforce integrity of user processes, Isabelle/ML requires user
  commands to be well-behaved in the above sense.  For example, output
  on raw \verb|TextIO.stdOut| from the Standard ML Basis Library
  results in a side-effect on that process channel that cannot be
  retracted by the transaction management of Isabelle.\footnote{This
  does not cause any further harm than potential user confusion, since
  unidentified output is redirected to some unmanaged console window,
  without any connection to the edited sources.}  In contrast, \verb|writeln| from the Isabelle/ML library attaches a message to the
  dedicated output stream of the running transaction; it will be
  located wrt.\ the original command span in the source text (within a
  certain document version), and disappear if the transaction is reset
  or discontinued due to document updates.

  \para{Implementation Notes.} The Isabelle/ML infrastructure to
  manage evaluation of user code has emerged over the last five years.
  Some of the main concepts are as follows.

  \begin{itemize}

  \item \emph{Unevaluated expressions} are represented by existing
  means of ML, either as unit abstraction \verb|fn () => a| of
  type \verb|unit -> 'a| or as regular function
  \verb|fn a => b| of type \verb|'a -> 'b|.  There are special
  combinators (variants of function application) that define a certain
  ``runtime mode'' for evaluation.  For example, the combinators \verb|uninterruptible| and \verb|interruptible| indicate that an ML
  expression is run with special thread attributes.

  \item \emph{Reified results} as explicit ML datatype that represents
  the disjoint sum of regular values or exceptional situations:

  \medskip
  \begin{tabular}{l}
  \verb|datatype 'a result = Res of 'a |\verb,|,\verb| Exn of exn| \\
  \verb|val capture: ('a -> 'b) -> 'a -> 'b result| \\
  \verb|val release: 'a result -> 'a| \\
  \end{tabular}

  The corresponding few lines of Isabelle/ML library help to organize
  evaluation of user code.  There are additional means to distinguish
  regular program exceptions from environmental effects (interrupts).

  Reified results are occasionally communicated explicitly in the
  interaction protocol.  For examples, a malformed Isabelle theory
  header in the editor buffer is already discovered as part of the
  organization of files in the front-end; it is passed through the
  protocol as \verb|Exn| value and produces a runtime error when
  the corresponding command transaction is run by the prover.  Thus we
  can formally hold up the requirement to make external syntax and
  protocol operations total, and postpone failures to the runtime
  environment within the prover.

  \item Functional wrappers for evaluation strategies as follows:

  \begin{itemize}

  \item Type \verb|'a future| represents value-oriented
  parallelism, with strict evaluation that is commenced eventually,
  unless the corresponding future task group is cancelled.  Regular
  results and program exceptions are memo-ized; environmental
  exceptions lead to an explicitly ``cancelled'' state of the future
  from which it cannot recover.  Future \emph{task} identifiers help
  to organize dependencies within the implicit queue, and hierarchic
  \emph{group} identifiers allow to define the propagation of
  exceptions and interrupts between peers and subgroup members.

  This task-parallel concept of Isabelle/ML is used to implement a
  small library of parallel list operations, with more conventional
  combinators like \verb|Par_List.map|, \verb|Par_List.exists| with full joining of results.

  \item Type \verb|'a promise|
  is a variant of \verb|'a future| without the built-in
  policies of parallel evaluation of closed expressions.  Instead,
  there is merely a synchronized single-assignment cell that is
  associated with a pro-forma future task, so that other future tasks
  can depend on it.  A promise can be fulfilled by external means, and
  thus cause other future evaluations to start.  This admits a
  form of reactive parallel programming in Isabelle/ML: open promises
  define the starting points of a dependency graph, with regular
  futures depending on them.  After the required promises are
  fulfilled, the parallel evaluation process starts to run, until
  completed or cancelled.

  \item Type \verb|'a lazy| represents expressions that are
  fully evaluated at most once, by an explicit force operation.
  Regular results and program exceptions are memo-ized, but not
  physical events.  Interrupting an attempt to force a lazy value will
  cause an interrupt of the caller, and reset the lazy value to its
  unevaluated state.  Note that multiple threads that happen to wait
  for a pending evaluation attempt might experience the interrupt.

  \item Type \verb|'a memo| is a synchronized single-assignment
  cell similar to \verb|'a lazy|, but \emph{with} memo-ization
  of interrupts.  In user-code, accidental absorption of physical
  events would lead to anomalies, but here we use it to organize
  incremental evaluation of user-code within the document.  After
  cancelling the current attempt to evaluate a document version, the
  system recovers from the partial result so far, and restarts any
  command transactions that have produced a result states with
  persistent interrupts.

  \end{itemize}

  \item External evaluation via some GNU bash script, to invoke
  arbitrary POSIX processes from the ML runtime environment, with
  propagation of interrupts in both directions.  For example, this
  allows to impose an ML-based timeout on external programs: the
  interrupt event that tells the corresponding ML thread to stop is
  turned into a process signal to terminate the POSIX process cleanly.

  \item Remote evaluation via an ML-Scala bridge, to invoke functions
  of type \verb|String => String| on the JVM.  Thus some ML
  worker thread temporarily transfers its runtime to a Scala
  counterpart.

  \end{itemize}

  Managed evaluation with different strategies is at the core of the
  Prover IDE concept.  It turns out as important prerequisite to
  advanced user-interaction, even before going into details
  of GUI programming.%
\end{isamarkuptext}%
\isamarkuptrue%
\isamarkupsubsection{Prover Output (PRINT) \label{sec:print}%
}
\isamarkuptrue%
\begin{isamarkuptext}%
The PRINT phase is somehow dual to READ (\secref{sec:read}).
  The original intention of the REPL model is to externalize the
  result of evaluation in a human-readable form, but this can mean
  many different things for proof assistants.

  Printing may already happen during evaluation, as a trace of the
  ongoing execution.  Conceptually, we decorate all prover messages
  with the \isa{id} of the running transaction, in order to
  re-assemble the output stream in the proper order, relatively to the
  original source of the command span or local positions within its
  source, say for warnings and errors that are directly attached to
  malformed ranges of source text.  The latter helps to reduce the
  non-determinism of command transactions that use parallel evaluation
  internally: results of different sub-evaluations usually refer to
  different parts of the source text.\footnote{Error messages that
  refer to the same source position are sorted according to the order
  they emerged within the prover.  The Isabelle/ML future library
  assigns serial numbers to ML exceptions internally.}

  Traditionally, the main result of an interactive proof step is the
  subsequent \emph{proof state}, which is printed implicitly for proof
  commands.  Since proof states often consist of large terms that
  require substantial time for printing (often more than the time for
  inferencing), it makes sense to organize the \isa{print} phase by
  continuing our REP diagram in the following manner:

  \medskip
  \begin{tabular}{lllllll}
  \isa{{\isasymdown}\isactrlsup r\isactrlsup e\isactrlsup a\isactrlsup d} &                & \isa{{\isasymdown}\isactrlsup r\isactrlsup e\isactrlsup a\isactrlsup d}   & & \isa{{\isasymdots}} \\
  \isa{st}  & \isa{{\isasymlongrightarrow}\isactrlsup e\isactrlsup v\isactrlsup a\isactrlisup l} & \isa{st{\isacharprime}}  & \isa{{\isasymlongrightarrow}\isactrlsup e\isactrlsup v\isactrlsup a\isactrlsup l} & \isa{st{\isacharprime}{\isacharprime}} & \isa{{\isasymdots}} \\
                &                 & \isa{{\isasymdown}\isactrlsup p\isactrlsup r\isactrlsup i\isactrlsup n\isactrlisup t} &               & \isa{{\isasymdown}\isactrlsup p\isactrlsup r\isactrlsup i\isactrlsup n\isactrlisup t} & & \isa{{\isasymdots}} \\
  \end{tabular}
  \medskip

  This means after the last \isa{eval} phase has finished, the
  system can fork the corresponding \isa{print} and proceed with the
  next \isa{eval}.  So the main evaluation thread will plough
  through the sequence of commands and fork many parallel \isa{print} tasks.  Doing this naively may saturate the future task queue
  with insignificant jobs that print proof states of invisible parts
  of the document, while the user is working elsewhere.

  This observation in earlier versions of our document model has
  motivated the notion of \emph{perspective}.  Thus the visible parts
  of the text are explicitly declared by suitable document edit
  operations, based on information from the physical editor and its
  views on the text buffer.  The \isa{print} phase is initialized as
  a lazy expression, which is turned into an active future only if the
  perspective uncovers it.  Afterwards it is guaranteed to finish,
  without any support to reset or cancel it.  This is sufficient under
  the assumption that printing always terminates in reasonable time.

  \medskip The above scheme integrates the traditional \isa{{\isasymPR}}
  command of Isabelle into the document model in a reasonably
  efficient manner.  If we consider long-running or non-terminating
  \isa{print} tasks not an accident, but a genuine concept to be
  supported, we could generalize \isa{{\isasymPR}} towards a whole class
  of diagnostic commands over finished command evaluations.  This
  would mean to piggy-back non-trivial analysis tools over prover
  commands, that analyze the situation and produce additional output
  for the user.  The existing portfolio of Isabelle tools like \isa{{\isasymNITPICK}}, \isa{{\isasymQUICKCHECK}}, or \isa{{\isasymSLEDGEHAMMER}}
  are examples for this.

  The implementation in Isabelle2013 still lacks this generalization
  of the \isa{print} phase towards arbitrary ``asynchronous agents''
  that interact with the document content after evaluation.  So far
  such functionality is simulated by inserting diagnostic commands
  into the document in the proper place, with the slight inconvenience
  of disrupting the evaluation of subsequent commands.%
\end{isamarkuptext}%
\isamarkuptrue%
\isamarkupsection{Protocol Interpreter \label{sec:protocol}%
}
\isamarkuptrue%
\begin{isamarkuptext}%
The classic REPL model makes a tight loop around the
  read-eval-print phases, to synchronize all phases immediately: emit
  a prompt and flush the output stream to re-synchronize with the
  input stream, and run the REP phases sequentially on the single
  main thread of the process.

  In contrast, our protocol interpreter that implements the
  document-oriented model (\secref{sec:document}) on the prover side
  works as follows.

  \begin{itemize}

  \item A dedicated \emph{protocol input thread} is connected to a
  private input channel from where it reads \emph{protocol commands},
  and evaluates them immediately.  This resembles some rudiments of
  the former REPL, but we merely do unidirectional stream processing,
  without re-synchronization by prompting the other side nor printing
  of results.

  \item Protocol commands are required to be \emph{total}, i.e.\ must
  not raise ML exceptions.  Error conditions need to be internalized
  into the protocol as separate messages returned to the front-end
  eventually.\footnote{Hard crashes of protocol commands are reported
  to a side-channel that is normally not shown to the user.}

  \item Protocol commands are required to terminate quickly, to keep
  the thread \emph{reactive}, e.g.\ within the range of 10--100\,ms.
  Note that the user-perception on the reactivity of the combination
  of editor front-end and prover back-end needs to take a full
  round-trip of certain protocol phases into account that are not
  explained in the present paper.

  \item Interrupts are blocked in the protocol thread; all operations
  on the main document state happen in a runtime context that is
  protected from physical events.  User events stemming from the
  editing process have already been \emph{internalized} as protocol
  commands into the stream of edits.  This also means that there is no
  longer any use of POSIX process signals, which were so hard to
  manage robustly and portably for multi-threaded processes.  User
  code is aborted exclusively via internal signals between ML threads,
  say as a consequence of cancellation of some future group by the
  protocol thread.

  \end{itemize}

  To make the protocol thread work reliably and efficiently, it is
  important to understand that protocol commands are not regular user
  commands.  The protocol defines a limited vocabulary of certain
  editing operations (such as \isa{Document{\isachardot}update} from
  \secref{sec:document}), which need to be applied in-place and
  reported to the front-end accordingly.  Prover commands occur as
  \emph{data} of such protocol commands, and are dispatched for
  independent evaluation on a separate thread farm of the future task
  library in Isabelle/ML.

  \para{Implementation Notes.} Early versions of the protocol
  interpreter imitated the classic Isar command loop by using \isa{stdin} and \isa{stdout} with quite concrete syntax for protocol
  commands and response messages, essentially an extension of the
  existing prover language with add-on commands like \isa{{\isasymUNDO}}
  or \isa{{\isasymREDO}} known from TTY mode.  This was adequate for
  prototypes, but had serious limitations in robustness and
  performance.

  For example, user-code may interfere with the global \isa{stdin{\isacharslash}stdout} streams of the process and disrupt the protocol.
  Classic TTY and Proof~General interaction is designed to tolerate
  this: the user can switch to the raw protocol buffer and insert some
  commands to recover from the confusion.  Such user intervention is
  no longer feasible in a system based on continuous streaming of
  document edits towards the prover, and results reported back from
  many transactions run in parallel.

  \medskip In the current production version the input channel is a
  private stream that is exclusively available to the protocol thread.
  On Unix we use named pipes (raw throughput \isa{{\isasymapprox}}~500\,MB/s)
  and on Windows the more portable TCP sockets (raw throughput \isa{{\isasymapprox}}~100\,MB/s).\footnote{Interestingly, much of this performance is
  lost due to the recoding of UTF-8 ML characters versus UTF-16 JVM
  characters.  This also explains why Isabelle/Scala sometimes prefers
  byte vectors that are presented as \texttt{CharSequence}, instead of
  the more bulky \texttt{String} type of Java.} Sockets require some
  effort to make them work in ML, but the Scala side consists only of
  a few lines of code.  In principle one could also run the protocol
  on a remote network connection (say via SSH tunneling), but the
  performance implications have not been explored yet.\footnote{The
  protocol uses relatively high band-width, but can afford long
  latency.  Current timeouts for flushing edits are in the comfortable
  range of 100--500\,ms, so one could try to reduce that to take
  network latency into account.  The protocol engine has been tested
  successfully with 0--1\,ms delay for its local buffers.}

  Spurious output on raw \isa{stdout{\isacharslash}stderr} is captured as well,
  and shown in a special console on demand.  Thus we handle tools
  gracefully that violate the official PRINT conventions
  (\secref{sec:print}).

  \medskip The protocol command syntax has been reduced to the bare
  minimum to maximize performance and robustness.  Errors in the
  encoding of the protocol would lead to failures that are difficult
  to repair, so we strive to avoid them by keeping it simple.

  Each command consists of a non-empty list of strings: name and
  arguments.  This structure is represented by explicit length
  indications in the protocol header, so that the protocol interpreter
  can read precise chunks from the stream without extra parsing.
  Decoding of arguments is left to the each protocol command
  implementation.

  There have been early attempts (inherited from PGIP) to use standard
  XML documents to carry protocol data, but it requires awkward
  maintenance of XML element names and XML attributes to accommodate
  the quasi-human-readable format and other complications of standard
  XML.  Instead, we now use a dedicated library in Isabelle/ML and
  Isabelle/Scala that performs data encoding of typed ML values over
  untyped XML trees, in the same manner as the ML compiler would do it
  for untyped bits in memory.  These raw XML trees are then
  transferred via YXML syntax \cite[\S2.3]{Wenzel:2011:CICM} in a
  robust way.

  This ML/XML/YXML data exchange is both efficient and easy to use,
  without demanding extra infrastructure for cross-language
  meta-programming (ML vs.\ Scala).  Runtime type-safety of such
  minimalistic marshalling of tuples, lists, algebraic datatypes etc.\
  is ensured by close inspection of a few lines of combinator
  expressions, both on the ML and the Scala side of the protocol
  implementation.  This untyped and unchecked approach works, because
  these program modules are maintained side-by-side in the same source
  code repository.  The accidental data formats that are encoded on
  the byte stream between the ML and Scala process are private to the
  implementation.  The public programming interface is defined by
  typed functions in ML or Scala, not the protocol messages
  themselves.%
\end{isamarkuptext}%
\isamarkuptrue%
\isamarkupsection{Conclusion%
}
\isamarkuptrue%
\begin{isamarkuptext}%
\begin{figure}[htb]
  \begin{center}
  \includegraphics[scale=0.4]{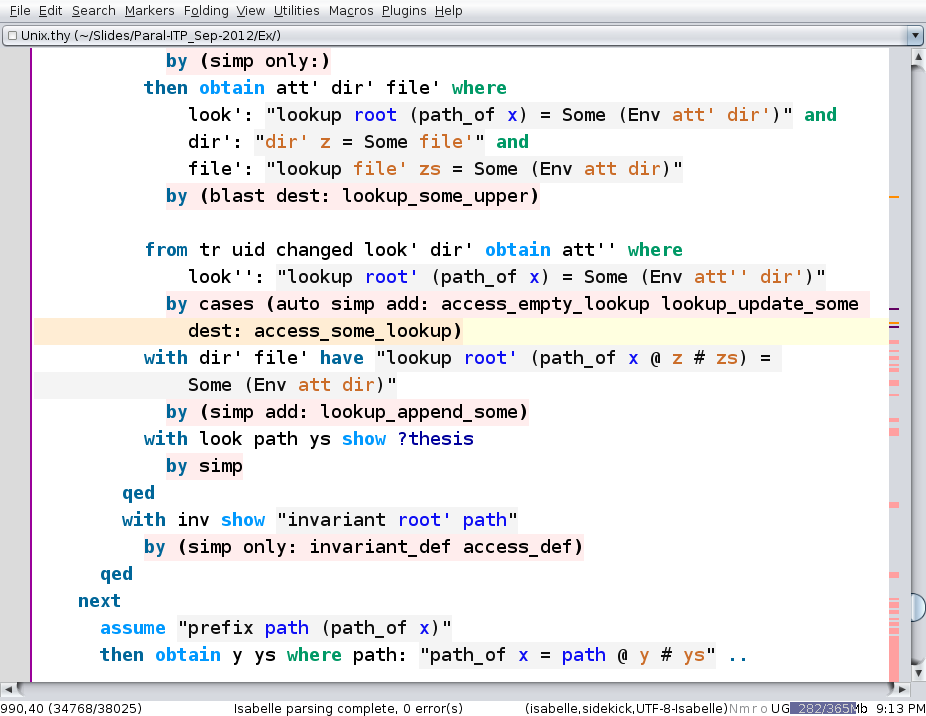}
  \end{center}

  {\small Status color code: pink = \emph{scheduled}, violet =
  \emph{running}, orange = \emph{finished with warning}, white/grey =
  \emph{finished}}

  \caption{Isabelle/jEdit with visualization of command status, both
  in the buffer and vertical status bar.}\label{fig:parallel_status}

  \end{figure}

  The issue of providing sophisticated user-interface support for
  sophisticated provers has been revisited many times over many years.
  Early efforts \cite{Bertot-Thery:1998} have eventually found their
  way into Proof~General \cite{Aspinall:TACAS:2000}, which is still
  the de-facto standard.  Its approach to wrap up the existing REPL of
  the prover has been continued by other projects like CoqIde
  \cite[\S4]{Wiedijk:2006} or Matita \cite{Asperti-et-al:2007}.

  The deeper reason for the success of the classic Proof~General
  approach is its conservativity wrt.\ the prover interaction model.
  Any prover that provides a reasonable REPL with some extra markup
  and an \isa{{\isasymUNDO}} command can participate.

  Investigating possibilities beyond Proof~General, Aspinall and
  others have already pointed out the need to reform provers
  themselves.  This eventually lead to the PGIP protocol definition
  \cite{Aspinall-et-al:2007} and its proposed front-end PG~Eclipse.
  The idea was to replace Emacs and Emacs LISP by industrial-strength
  Eclipse and Java.  In retrospective, there are a variety of reasons
  why this approach did not become popular in the proof assistant
  community: it demands substantial efforts to implement and maintain
  PGIP on the prover side, and requires people who are accustomed to
  think in terms of higher-order logic and dependent types to engage
  in profane Java and XML.  Moreover, the interaction model of PGIP is
  still quite close to classic Proof~General, so the returns for the
  investment to support it were not sufficient.

  \medskip Our strategy to bridge the cultural gap between ML and the
  JVM is based on Scala \cite{Scala:2004}.  After substantial reforms
  on the prover side, the current state of concepts and implementation
  of Isabelle/ML/Scala and Isabelle/jEdit as Prover IDE on top of it
  have reached a state where other projects can join the effort,
  either on the back-end or front-end side.  Thus we hope to reform
  and renovate of the LCF-approach to interactive theorem proving for
  coming decades of applications.  There is ongoing work with some Coq
  developers to transfer some of the ideas presented here to their
  particular prover \cite{Tassi-Barras:2012}.

  The existing Isabelle implementation is expected to improve further
  in the near future.  The main conceptual omission is the management
  of diagnostic commands over proof documents (cf.\ the discussion of
  the PRINT phase).  We intend to support a notion of ``asynchronous
  agents'' natively, which will allow to attach automated provers and
  disprovers provided via Sledgehammer and Quickcheck.  Such advanced
  modes of tool-assisted proof authoring need to be worked out
  further and turned into practice.

  \medskip Ultimately, the goal is to improve performance and
  reactivity of prover interaction, especially on multi-core hardware
  \cite{CICM:2013:Paral-ITP}.  The present paper concentrates on the
  impact on the prover architecture.  In order to assemble the final
  Prover IDE, the prover front-end technology needs to taken into
  account as well.

  For example, consider the vertical command status bar in
  Isabelle/jEdit (Figure~\ref{fig:parallel_status}), at the right of
  the regular text view.  It visualizes the status information of all
  commands within a theory node, according to the generalized REP
  concepts discussed in this paper.  Retrieving that information for
  the whole theory lacks the locality of the main editor view on the
  text: beyond 100\,kB theory size the time to repaint the side-bar
  exceeds 50\,ms and thus the editor becomes inconvenient to use:
  painting is synchronous in the central GUI thread of Java/Swing, so
  it inhibits further user input for fractions of a second.  Fully
  asynchronous and parallel GUI operations are difficult to achieve in
  Java/Swing applications: the visualization of massive amounts of
  data from the parallel prover requires further investigation in the
  future.  In Isabelle2013 this problem is circumvented by
  restricting the logical scope of that status bar to a certain
  amount of text, as specified via user preferences.%
\end{isamarkuptext}%
\isamarkuptrue%
\isadelimtheory
\endisadelimtheory
\isatagtheory
\isacommand{end}\isamarkupfalse%
\endisatagtheory
{\isafoldtheory}%
\isadelimtheory
\endisadelimtheory
\isanewline
\end{isabellebody}%

\bibliographystyle{eptcs}
\bibliography{root}

\end{document}